\documentstyle[prl,aps,epsfig,multicol]{revtex}
\begin{document}
\draft

\hfuzz=100pt
\hbadness=10000


\title{Punctuated evolution for the quasispecies model}

\author{Joachim Krug and Christian Karl}
\address{Fachbereich Physik, Universit\"at Essen, 45117 Essen, Germany}

\date{\today}
\maketitle

\begin{abstract}
Biological evolution in a sequence space with random fitnesses is studied within
Eigen's quasispecies model. A strong selection limit is employed, in 
which the population resides at a single sequence at all times.
Evolutionary trajectories start at a randomly chosen sequence and proceed
to the global fitness maximum through a small number of intermittent jumps.
The distribution of the total evolution time displays a universal power law
tail with exponent -2. Simulations show that the evolutionary 
dynamics is very well represented by a simplified shell model, in which
the sub-populations at local fitness maxima grow independently. The shell
model allows for highly efficient simulations, and provides a 
simple geometric picture of the evolutionary trajectories.

\end{abstract}


\vskip2pc
\begin{multicols}{2}
\narrowtext

Biological evolution often displays a \emph{punctuated} 
dynamical pattern, in the sense that quiescent periods of \emph{stasis}
alternate with bursts of rapid change. A variety of mechanisms for
punctuation have been proposed, which operate on different levels
of the tree of life. On the largest scales of macroevolution, 
coevolutionary avalanches may play a role, which have been associated with
self-organized criticality \cite{Bak93}. On the level of populations,
punctuation due to rare, beneficial mutations has been observed
in evolution experiments with bacteria \cite{Elena96}. Similar behavior
has been found in simulations of RNA evolution, where stasis
corresponds to diffusion on a neutral network, and a punctuation event 
marks the transition to another network of higher fitness \cite{Fontana98}.

Possibly the simplest interpretation of punctuated evolution is in terms of 
a homogeneous population, represented by a localized distribution
in some phenotypic
or genotypic space, which evolves in a static, multipeaked fitness
landscape \cite{Simpson44}. Under conditions of strong selection and
small mutation rate, such a population will rapidly climb a local fitness
maximum, where it then resides for a long time, until a rare, large fluctuation
allows it to cross the valley to a more favorable peak. At least in the
limit of infinite population size \cite{VanNimwegen00}, the mathematics
of this process is closely related to physical problems such as noise-driven
barrier crossing, tunneling \cite{Ebeling84}
 and variable-range hopping \cite{Zhang86}, 
and it is easy to show that the residence time at one peak can be vastly larger than the
time required for the transition to the next \cite{Newman85}.
In a rugged fitness landscape, the sequence of transitions forms an 
\emph{evolutionary trajectory}, which probes the distribution of
fitness peaks and the geometry of the landscape. 

In this Letter we investigate the statistics of such evolutionary
trajectories in the framework of Eigen's quasispecies model 
\cite{Eigen71,Eigen89}. We consider a population of 
individuals, each characterized by a binary genomic sequence $\sigma$
of length $N$, which reproduce asexually 
and mutate in discrete time $t$.
The total number of sequences is $S = 2^N$.
An individual with genotype $\sigma$ leaves $A(\sigma)$ offspring
in the next generation, and point mutations occur with probability
$\mu$ per site and generation. In a mean field approximation,
which neglects finite population effects as well as nonlinear
constraints on the population size, this leads to the evolution
equations 
\begin{equation}
\label{rate}
z(\sigma,t+1) = \sum_{\sigma'} A(\sigma') \mu^{d(\sigma,\sigma')}
(1 - \mu)^{N-d(\sigma,\sigma')} z(\sigma',t)
\end{equation}
for the number $z(\sigma,t)$ of individuals of genotype $\sigma$,
where $d(\sigma,\sigma')$ denotes the Hamming distance between
sequences $\sigma$ and $\sigma'$, i.e. the number of symbols in which
the two differ. 

The fitness landscape enters (\ref{rate}) through the choice
of $A(\sigma)$. For single peaked landscapes, the model is known
to exhibit the \emph{error threshold} phenomenon, in which a
population which is concentrated around the fitness peak -- the
\emph{quasispecies} -- becomes delocalized with increasing mutation rate,
increasing sequence length, 
or decreasing peak height \cite{Eigen71,Eigen89,Galluccio97}. 
Here we want to study the sudden transitions between the different quasispecies
associated with the multiple 
peaks of a rugged fitness landscape \cite{McCaskill84}.
This punctuated behavior becomes more pronounced the more strongly 
the peaks are able to localize the population. We therefore pass
to a \emph{strong selection limit} \cite{Krug01}, which is motivated
by the zero temperature limit of the equivalent problem of a pinned
directed polymer \cite{Galluccio97,Krug93}. To this end the reproduction 
rates are written as $A(\sigma) = e^{k F(\sigma)}$, where 
$k$ is an \emph{inverse selective temperature} \cite{Peliti97}. 
In the limit $k \to \infty$ the population variables
take the form $z(\sigma,t) = e^{k E(\sigma,t)}$ and the mutation
rate per generation has to be scaled down as $\mu = e^{-\gamma k}$ for
some constant $\gamma > 0$. Then (\ref{rate}) reduces to 
\begin{equation}
\label{strong}
E(\sigma,t+1) = \max_{\sigma'}
[E(\sigma',t) + F(\sigma') - \gamma d(\sigma,\sigma') ].
\end{equation}
The fitnesses $F(\sigma) \geq 0$ are chosen
randomly and independently from 
a distribution $p(F)$. The numerical simulations presented in this
paper were performed with $p(F) = e^{-F}$, but results for other
distributions will be described as well.
The value of $\gamma$ is set to unity. 

Evolutionary trajectories are generated as follows. At time $t=0$
the population is placed at a randomly chosen 
sequence $\sigma^{(i)}$, corresponding to the initial condition 
$E(\sigma^{(i)},0) = 0$, $E(\sigma \neq \sigma^{(i)} ,0) = - \infty$.
Subsequently (\ref{strong}) is iterated, and the position 
$\sigma^\ast(t)$ of the
global maximum of $E(\sigma,t)$ is recorded; in the strong selection
limit, essentially the whole population resides at $\sigma^\ast(t)$
at time $t$. The trajectory $\sigma^\ast(t)$ diplays the expected
punctuated pattern \cite{Krug01}. After the first
few time steps, it passes exclusively through local fitness maxima.
The transitions between different local maxima are abrupt, and the 
time period spent at a given peak, as well as the probability for
a peak to be visited by a trajectory
with random starting point, increases strongly with its fitness.
The end point of each trajectory is the global fitness 
maximum of the landscape.

\begin{figure}
\centerline{\epsfig{figure=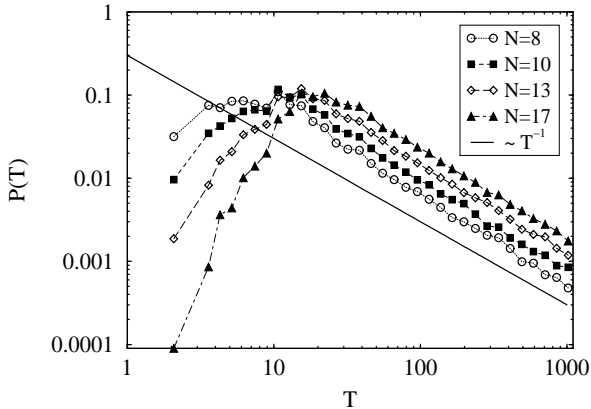,height=8cm,width=6cm,angle=-90}}
\vspace*{0.5cm}
\caption{Distribution of evolution times for the local model. 
The data have been binned
exponentially, hence the measured exponent is $\alpha - 1$.}
\label{PT-fig}
\end{figure}

\begin{figure}
\centerline{\epsfig{figure=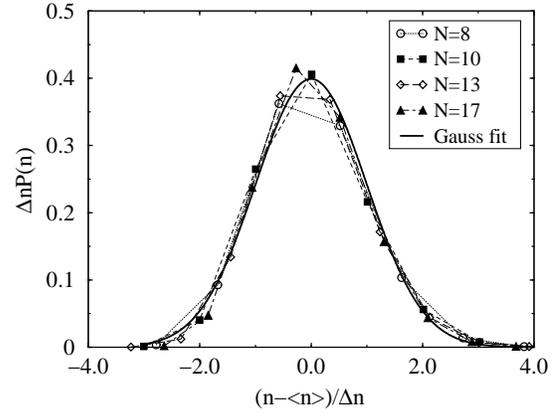,height=8cm,width=6cm,angle=-90}}
\vspace*{0.5cm}
\caption{Distribution of the number of jumps for the local model,
shifted and scaled by its mean and standard deviation.}
\label{Pn-fig}
\end{figure}

Here we focus on two statistical measures of the evolutionary
trajectories: The distribution $P(T)$ of times $T$ required to reach
the global maximum, and the distribution $P(n)$ of the number $n$ of
jumps between local maxima that occur on the way. In Figs.\ref{PT-fig}
and \ref{Pn-fig} numerical results for sequence lengths $N = 8$ -- 17
are shown. The data were averaged over $10^5$ trajectories, each
evolving in a separate, independently generated
fitness landscape. 
To make the problem numerically tractable despite the
exponential growth of sequence space with increasing $N$, these
simulations were carried out using a \emph{local} version of 
the recursion relation (\ref{strong}), in which the maximization
on the right hand side is performed over nearest neighbor sequences
with $d(\sigma,\sigma') \leq 1$ only. This implies that at most 
one point mutation is allowed per generation. 
The full dynamics (\ref{strong}) and the local version 
will be referred to as the global and the local model, respectively.
The differences in behavior that arise due to the local restriction
will be explained below. 

Notable features of the numerical results in Fig.\ref{PT-fig} include
(i) a (roughly linear) increase of the \emph{typical} evolution time
with $N$,
as represented e.g. by the maximum of $P(T)$, and (ii) a power law
tail $P(T) \sim a(N) T^{-\alpha}$ with $\alpha \approx 2$ for $T \gg N$. 
A careful analysis of the data for different sequence lengths shows
no systematic dependence of $\alpha$ on $N$, and yields the 
overall best estimate $\alpha = 2.13 \pm 0.04$. 
If it is assumed that $\alpha = 2$ exactly (see below), then the
coefficient $a(N)$ of the power law scales with sequence length
as $a(N) \sim N^{1.7}$. 
The distribution of the number of jumps in 
Fig.\ref{Pn-fig} is well fitted by a Gaussian. The mean number of 
jumps $\bar n$ is small, and the 
variance $\Delta n^2 < {\bar n}$, indicating sub-Poissonian 
fluctuations. In the (admittedly small) range of $N$
accessible to our simulations, the dependence on 
sequence length is reasonably
well described by power laws, ${\bar n} \approx 0.56 \cdot
N^{0.73}$ and
$\Delta n^2 \approx 0.17 \cdot N^{0.74}$.

To gain some insight into these results, we now introduce
an approximation to (\ref{strong}) which allows for some 
analysis, as well as for a much more efficient numerical 
scheme. The evolutionary dynamics 
(\ref{rate}) and (\ref{strong}) involves two distinct processes:
The \emph{spreading} of the population through sequence space by mutations,
and the (exponential) \emph{growth} of the local sub-populations by selection.
Our key assumption is that the spreading part of the dynamics is 
important only in the initial stages of evolution, while the late
stage can be described as a competition between independently
growing quasispecies associated with different local fitness
maxima. 

\begin{figure}
\centerline{\epsfig{figure=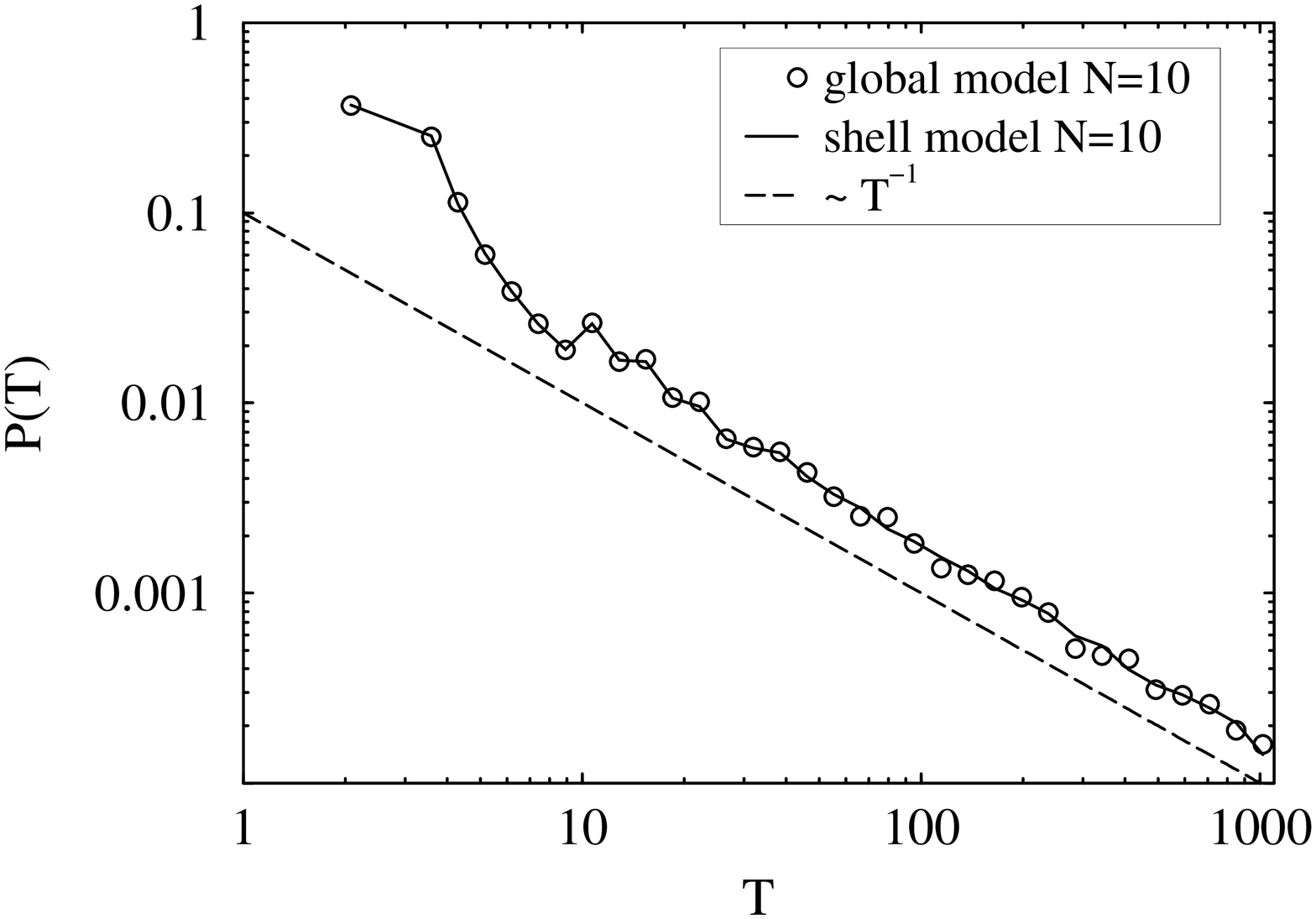,height=8cm,width=6cm,angle=-90}}
\centerline{\epsfig{figure=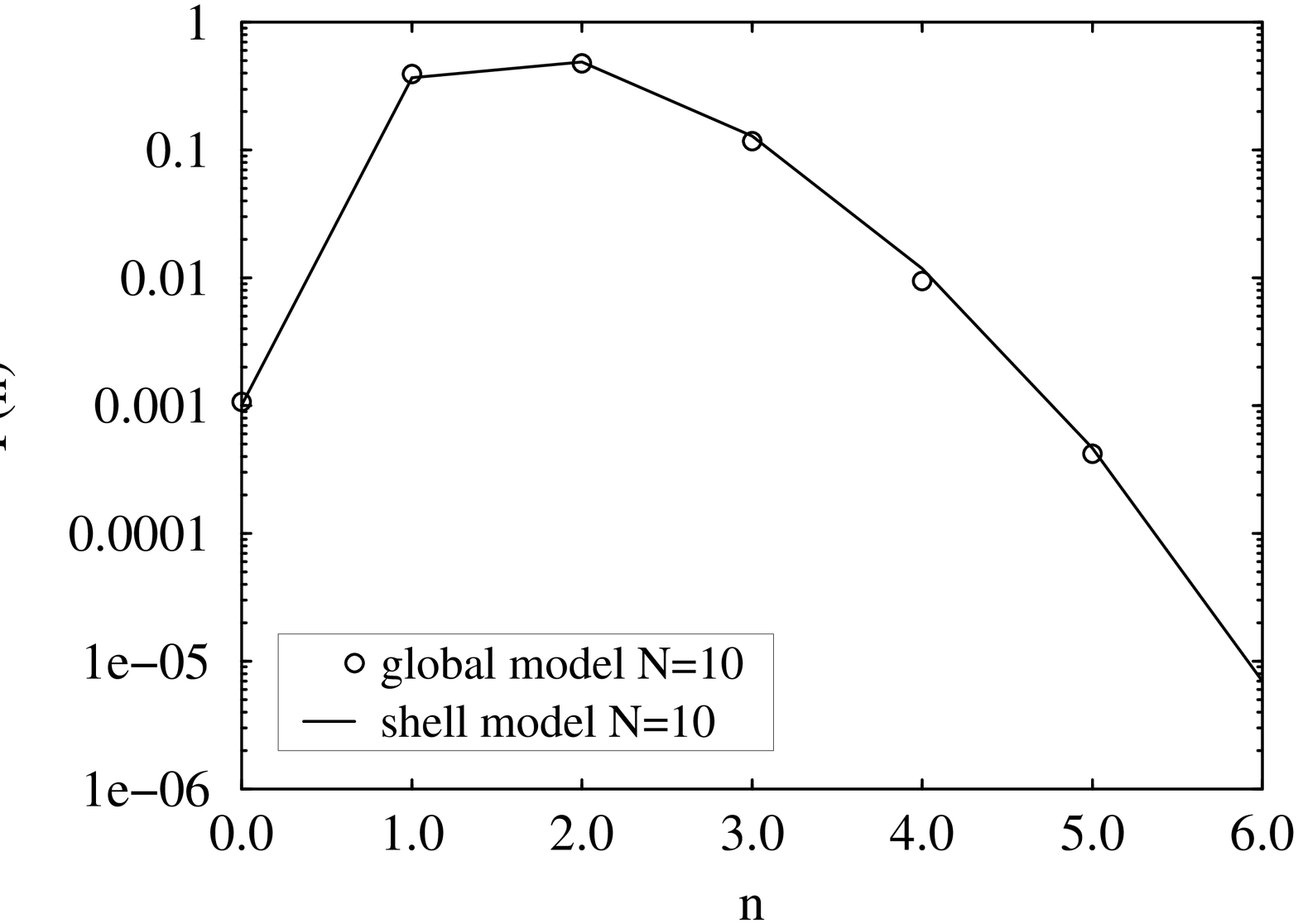,height=8cm,width=6cm,angle=-90}}
\vspace*{0.5cm}
\caption{Comparison between the full global dynamics
and the shell model for sequence length $N=10$. The number of realizations
was $10^6$ for the full model and $10^7$ for the shell model. Upper panel: 
Binned distribution of evolution times. Lower panel: Distribution of the 
number of jumps.
}
\label{Comp-fig}
\end{figure}

More specifically, after the first time step
the recursion relation (\ref{strong}) with the initial
condition localized at $\sigma^{(i)}$ yields a 
(logarithmic) population distribution
$E(\sigma,1) = F(\sigma^{(i)}) - \gamma d(\sigma,
\sigma^{(i)})$ which is symmetric around $\sigma^{(i)}$ 
and linearly decreasing with increasing distance from
$\sigma^{(i)}$. Thus at $t=1$ every sequence is \emph{seeded} with
a (possibly astronomically small) population. 
The approximation consists of letting these populations evolve
independently for $t \geq 2$, i.e. to replace (\ref{strong})
by the simple linear growth law
\begin{equation}
\label{linear} 
E(\sigma,t) = E(\sigma,1) + (t-1) F(\sigma).
\end{equation} 
Since $E(\sigma,1)$ is constant on the \emph{shells} of constant
Hamming distance $d(\sigma,\sigma^{(i)}) = k$ to the initial population,
for each shell only the sequence with the largest fitness needs
to be taken into account. The $N$ shell fitnesses can be generated
directly from the distribution of the largest among ${N \choose k}$
exponential random variables, corresponding to the ${N \choose k}$ 
sequences contained in the shell. In this way the shell
approximation vastly reduces the computational effort from 
$2^N$ to $N$. Fig.\ref{Comp-fig} shows that the shell approximation works
astoundingly well. For sequence lengths $N=8$ and 10, for which a comparison
is possible,  
the distributions $P(T)$ 
and $P(n)$ obtained (with little effort) using the shell approximation
are indistinguishable from those generated by the full model 
(\ref{strong}) with global maximization.

The situation is more
complicated for the local
model used in Figs.\ref{PT-fig} and \ref{Pn-fig}, because
in that case the seeding stage takes $N$ time steps. This explains
immediately the appearance of a typical evolution time of order 
$N$, which is absent in the global model: On average, it takes
$N/2$ time steps before the global fitness maximum has even been seeded.
The seeding time for the shell at distance $d(\sigma,\sigma^{(i)}) = k$
is $k$. 
The shell approximation can be implemented only for times $t \geq k$,
where linear growth according to 
$
E(\sigma,t) = E(\sigma,k) + (t - k) F(\sigma)
$
sets in. This is however less useful than (\ref{linear}), because
the computation of $E(\sigma,k)$ already requires the solution of 
a nontrivial optimization problem: The population strength at 
the ``seeding front'' defined by $d(\sigma,\sigma^{(i)}) = t$ is given by
\begin{equation}
\label{front}
E(\sigma,t) = \max_{\pi}
\left[\sum_{\sigma' \in \pi} F(\sigma') \right] - \gamma t,
\end{equation}
where $\pi$ denotes a directed path of length $t$ from 
$\sigma^{(i)}$ to $\sigma$. 
A reasonable shell approximation for the local model
(which is however not as accurate as in the global case)
is obtained by treating the first term
on the right hand side of (\ref{front}) as a constant,
setting $E(\sigma,k) = - \gamma k$ in shell $k$. 

Using (\ref{linear}), the total evolution time $T$ can be 
estimated as 
\begin{equation}
\label{T}
T \approx \frac{\gamma[d(\sigma^{(f)},\sigma^{(i)}) - 
d(\sigma^{(f-1)},\sigma^{(i)})]}{F(\sigma^{(f)}) - F(\sigma^{(f-1)})},
\end{equation}
where $\sigma^{(f)}$ denotes the global fitness maximum, and 
$\sigma^{(f-1)}$ is the last sequence visited before the global
maximum is reached. We expect that $\sigma^{(f-1)}$ is close in fitness
to the sequence with the second largest fitness in the system, and
hence the denominator in (\ref{T}) is comparable to the 
\emph{fitness gap} $\epsilon$ of the landscape, which we define as
the difference between the largest and the second largest of the 
$S$ fitness values. To estimate the numerator of (\ref{T}), we
note that most sequences reside in a belt of thickness $\sim \sqrt{N}$
around the mean distance $N/2$ from $\sigma^{(i)}$, and hence
$d(\sigma^{(f)},\sigma^{(i)}) - d(\sigma^{(f-1)},\sigma^{(i)}) \sim
\sqrt{N}$. We conclude that $T \sim \sqrt{N}/\epsilon$. The inverse
correlation between $T$ and $\epsilon$ for a given landscape has been
explicitly verified in the simulations \cite{Karl01}. 

The distribution $P_g(\epsilon)$
of the fitness gap can be computed from order
statistics \cite{David70}. 
The distribution of evolution times then becomes
$P(T) \sim (\sqrt{N}/T^2) P_g(\sqrt{N}/T)$, which displays a power
law tail with exponent $\alpha = 2$ for $T \gg \sqrt{N}$
and prefactor $a(N) \sim \sqrt{N} P_g(0)$. 
The $T^{-2}$-decay is universal, because $P_g(0)$ is always finite
and nonzero \cite{Krug01}. It follows from the explicit expression 
that
$P_g(0) \sim (\Delta F_{\mathrm{max}}(S))^{-1}$, where 
$\Delta F_{\mathrm{max}}(S)$ denotes the standard deviation of the maximum
$F_{\mathrm{max}}(S)$
among the $S$ independent fitness values. 
For the exponential fitness distribution
$P_g(\epsilon) = \exp(-\epsilon)$  \emph{independent}
of $N$, so that $a(N) \sim \sqrt{N}$. In contrast, for fat tailed
fitness distributions such as power laws, $a(N)$
\emph{decreases} with increasing $N$. These predictions
are well confirmed by simulations of the
global shell model. 

Repeating the argument for the
shell approximation to the local model described above, 
one finds 
$P(T) \sim a(N) T^{-2}$ with
$a(N) \sim \sqrt{N} \langle F_{\mathrm{max}}(S) \rangle
P_g(0)$.
For the exponential distribution 
$\langle F_{\mathrm{max}}(S) \rangle = \ln S \sim N$,
hence $a(N) \sim N^{3/2}$,
which is consistent with the numerical estimate quoted above.
We conclude that the asymptotic value of the evolution time exponent
is $\alpha = 2$ in all cases, and attribute the deviations found
in the data in Fig.\ref{PT-fig} to short time
corrections originating in the seeding stage.

Within the shell model, the problem of determining the number of 
jumps along an evolutionary trajectory has a simple geometric
interpretation. Eq.(\ref{linear}) defines a family of $N$ straight
lines in the $(t,E)$-plane, one for each shell. 
The intercept of the line
corresponding to shell $k$ is $F(\sigma^{(i)}) - \gamma k$ and its
slope is the shell fitness $F_k$, i.e. the largest
among ${N \choose k}$ independent fitness values. The evolutionary
trajectory $\sigma^\ast(t)$ then corresponds to the upper envelope
of this family of lines, and the number of jumps is the 
\emph{number of corners} of this random polygon. 
It follows immediately that the number of jumps is always smaller
than $N$. To go beyond this trivial bound is not easy, because
the shell fitnesses are not identically distributed, i.e. the distribution
of $F_k$ depends on $k$. For large $N$ the
global maximum is certain to lie near the shell $k=N/2$,
so that $n \leq N/2$. If one assumes that only
sequences with fitnesses comparable to the the global maximum, which
reside in the belt around $k = N/2$, contribute to the trajectory,
then $\bar n \sim \sqrt{N}$. Simulations
of the global shell model yield $\bar n \sim N^{0.60}$
and $\Delta n^2 \sim N^{0.72}$
for $4 \leq N \leq 29$, but the data are not inconsistent with
a linear dependence on $\sqrt{N}$.

More progress is possible for a simplified shell model in which
the shell fitnesses are identically distributed (this corresponds
to a one-dimensional sequence space). Suppose that at some time $t$
the population resides in shell $k$. The next jump then occurs to 
the shell $k' > k$ for which (i) $F_k' > 
F_k$ and (ii) the time 
$
t(k,k') = \gamma(k' - k)/(F_k' - F_k)
$
at which the two lines cross is minimal among all such $k'$. If the 
second requirement is ignored and one simply choses the next shell
with $k' > k$ and $F_k' > 
F_k$, the problem reduces to that of record dynamics,
for which the number of jumps is Poisson distributed with mean
$\ln N$ \cite{Sibani}. Since the greedy record algorithm evidently
produces more jumps than the original dynamics, 
the mean number of jumps in the uniform shell model is bounded
from above by $\ln N$, independent of the fitness distribution.
Simulations of this model 
indicate that in general $\bar n \approx \beta \ln N$, where
the coefficient $\beta < 1$ depends on the extremal statistics of the
fitness distribution. Specifically, we conjecture that 
$\beta = 1/2$ for distributions similar to an exponential,
$\beta = (\delta - 1)/(2 \delta - 1)$ for power law distributions
$p(F) \sim F^{-(\delta + 1)}$, and $\beta = (2 + \nu)/(3 + 2 \nu)$
for bounded distributions with $p(F) \sim (F_0 - F)^{\nu}$, 
$F \leq F_0$. 

In conclusion, we have explored the statistical 
properties of evolutionary trajectories in a 
sequence space equipped with a rugged fitness landscape.
The simple but highly accurate shell approximation elucidates
the interplay of the geometry of sequence space and the
extremal statistics of fitness values, and allows for 
efficient numerical investigations. It seems worthwhile
to extend these studies to correlated fitness landscapes and
different types of graphs, with the aim of using evolutionary
trajectories as a means to characterize the geometry of general
optimization problems.

Useful discussions with H. Flyvbjerg, T. Halpin-Healy, L. Peliti
and K. Sneppen are gratefully acknowledged. We have enjoyed the
hospitality of Barnard College (New York), NBI (Copenhagen), DTU
(Lyngby), ITP (Santa Barbara) and SNBNCBS (Kolkata) 
during various stages of this work. 
Support has been provided by NATO within CRG.960662,
DFG within SFB 237 and by NSF under Grant No. PHY99-07949.


\end{multicols}

\end{document}